# Spoof Plasmon Hybridization

Jingjing Zhang[2,*], Zhen Liao[1,*], Yu Luo[2,*,#], Xiaopeng Shen[1], Stefan A. Maier[3,#], Tie Jun Cui[1,#]

[1] State Key Laboratory of Millimetre Waves, Southeast University, Nanjing 210096, China

[2] School of Electrical and Electronic Engineering, Nanyang Technological University, Nanyang Avenue 639798, Singapore

[3] The Blackett Laboratory, Department of Physics, Imperial College London, London SW7 2AZ, UK

* These authors contributed equally to this work.

# To whom correspondence should be addressed.

Emails: luoyu@ntu.edu.sg; s.maier@imperial.ac.uk; tjcui@seu.edu.cn

## Abstract

Plasmon hybridization between closely spaced nanoparticles yields new hybrid modes not found in individual constituents, allowing for the engineering of resonance properties and field enhancement capabilities of metallic nanostructure. Experimental verifications of plasmon hybridization have been thus far mostly limited to optical frequencies, as metals cannot support surface plasmons at longer wavelengths. Here, we introduce the concept of 'spoof plasmon hybridization' in highly conductive metal structures and investigate experimentally the interaction of localized surface plasmon resonances (LSPR) in adjacent metal disks corrugated with subwavelength spiral patterns. We show that the hybridization results in the splitting of spoof plasmon modes into bonding and antibonding resonances analogous to molecular orbital rule and plasmonic hybridization in optical spectrum. These hybrid modes can be manipulated to produce enormous field enhancements (>5000) by tuning the separation between disks or alternatively, the disk size, which effectively changes the relative gap size. The impact of the radiation loss is considered to find out the optimum disk size that maximizes field enhancement capabilities. Our investigation not only extends the range of applicability of the hybridization model, but also provides insightful guidance to exporting the exciting applications associated with plasmon hybridization to lower spectral range.

# 1. Introduction

Metallic nanoparticles support surface plasmons at optical frequencies and have the ability to concentrate electromagnetic radiation into nanoscale dimensions, giving rise to a strong enhancement of electromagnetic fields in the near-field region. In complex plasmonic systems consisting of multiple closely spaced nanoparticles, the localized surface plasmon resonance (LSPR) modes of each individual nanoparticle can interact and form hybridized bonding and antibonding plasmon modes. Such coupling effect allows for largely increased field enhancement at LSPR compared to isolated constituents, making it attractive for a large variety of applications, including molecular sensing [1], surface-enhanced spectroscopy [2], photovoltaics [3], and nonlinear process [4]. As the geometrical features of the nanoparticles are pushed to sub-nanometer regime, nonlocal and quantum effects arising from the dispersion of charges at the metal boundaries play increasingly important role, setting an ultimate limit for the maximum field enhancement [5-10]. At lower frequencies (terahertz regions or microwave frequencies) where the effects of metal smearing are negligible, however, metals feature poor confinement of electromagnetic waves and cannot support surface plasmons. Recently, by patterning the metal surface with subwavelength periodic features, spoof surface plasmon polaritons (SPPs) have been developed from microwave to far-infrared frequencies to control the surface electromagnetic modes with properties resembling those of conventional metallo-dielectric plasmonic structures at optical frequencies [11-15]. This metamaterial approach offers possibilities of bringing most of the advantages associated with optical SPPs to lower frequencies, such as high confinement of electromagnetic field [13]. In particular, in the terahertz and microwave spectra where the metal is highly conductive, even higher field enhancements may be achieved with particles supporting spoof SPPs, as compared to their optical counterparts. Moreover, the behaviour of spoof SPPs depends primarily on the geometry of the corrugations instead of the optical properties of the metal. This gives us extended flexibility in controlling the surface plasmons, either as propagating SPPs in the waveguide configurations [14-19] or as localized surface plasmons

(LSPs) in structures with closed surfaces [20-25]. While various theoretical and experimental studies have been performed on the spoof SPPs in individual devices or structures, the investigation on the hybridization of spoof LSP modes in coupling systems have not yet been reported.

A thorough understanding of the coupling between individual structures can facilitate the design of highly sophisticated plasmonic systems with desired optical properties. An electromagnetic analogue of the molecular orbital theory, the plasmon hybridization model has been proposed as a simple and intuitive tool to describe and interpret the plasmon response of complex nanostructures or interacting nanoparticle systems. In this work, we present a hybridization scheme of spoof SPPs, showing that the spoof LSP resonance properties of the complex structures depends on the geometries and spectra of the building blocks. In analogy to the rule of molecular orbitals or plasmonic hybridization, the collective modes supported by such spoof plasmonic system can be categorized into bonding and antibonding resonances [26]. In particular, we investigate experimentally the coupling between adjacent metal disks with spiral subwavelength textures which support spoof LSP at microwave frequencies. We show that the hybrid spoof LSPs can be manipulated to produce enormous field enhancement by tuning the separation in the dimer or alternatively, the radius of the disk, which effectively changes the relative gap size. The interplay between the radiative loss and the plasmonic focusing is investigated in details, and the optimum size parameter is found for which the field enhancement is maximized. Our investigation not only extends the range of applicability of the hybridization model, but also provides insightful guidance to exporting the exciting applications associated with plasmon hybridization to lower spectral range.

## 2. Hybridization Model for Spoof Plasmons

A core-shell nanostructure represents a simple prototypical model system for the study of plasmon hybridization, where the geometry-dependent plasmon response can be taken as an interaction between the plasmon resonances of a nanosphere and a nanocavity [26], resulting in the splitting of the plasmon resonances into a lower energy symmetric (bonding) mode and a higher energy antisymmetric (antibonding) mode. Here, we consider a two-dimensional (2D) perfect electric conductor (PEC) composite structure comprising a cavity with spiral corrugations (with inner radius $r_2$, outer radius $R_2$, and spiral intersection angle $\theta_2$) and a spiral core (with inner radius $r_1$, outer radius $R_1$, and spiral intersection angle $\theta_1$) as illustrated in Fig. 1 (a). For the cases where the ratio of the spacing between the spiral arms to the period of the spiral is fixed as $\frac{a}{d}$, the surface-charge density can be expressed as cylindrical harmonics of order $n$, and the resonances modes can be derived as (Details given in the supporting information)

$$\left(\frac{r_2}{R_1}\right)^{2n} = \frac{\left(1 - \frac{k_0 a R_1 \cos\theta_1 \alpha}{n n_g d}\right)\left(1 - \frac{k_0 a r_2 \cos\theta_2 \beta}{n n_g d}\right)}{\left(1 + \frac{k_0 a R_1 \cos\theta_1 \alpha}{n n_g d}\right)\left(1 + \frac{k_0 a r_2 \cos\theta_2 \beta}{n n_g d}\right)} \quad (1)$$

where $n_g$ is the dielectric constant of the surrounding medium, $k_0$ is the free space wave vector; $\alpha$ and $\beta$ are defined as

$$\alpha = \frac{J_1\left(\frac{n_g k_0 r_1}{\cos\theta_1}\right) Y_1\left(\frac{n_g k_0 R_1}{\cos\theta_1}\right) - J_1\left(\frac{n_g k_0 R_1}{\cos\theta_1}\right) Y_1\left(\frac{n_g k_0 r_1}{\cos\theta_1}\right)}{J_0\left(\frac{n_g k_0 R_1}{\cos\theta_1}\right) Y_1\left(\frac{n_g k_0 r_1}{\cos\theta_1}\right) - J_1\left(\frac{n_g k_0 r_1}{\cos\theta_1}\right) Y_0\left(\frac{n_g k_0 R_1}{\cos\theta_1}\right)},$$

and

$$\beta = \frac{J_1\left(\frac{n_g k_0 r_2}{\cos\theta_2}\right) Y_1\left(\frac{n_g k_0 R_2}{\cos\theta_2}\right) - J_1\left(\frac{n_g k_0 R_2}{\cos\theta_2}\right) Y_1\left(\frac{n_g k_0 r_2}{\cos\theta_2}\right)}{J_0\left(\frac{n_g k_0 r_2}{\cos\theta_2}\right) Y_1\left(\frac{n_g k_0 R_2}{\cos\theta_2}\right) - J_1\left(\frac{n_g k_0 R_2}{\cos\theta_2}\right) Y_0\left(\frac{n_g k_0 r_2}{\cos\theta_2}\right)}.$$

Note that this expression for eigenmodes is similar to the case of metallic or dielectric core-shell nanostructure where the resonance condition is expressed as [26, 27]

$$\left(\frac{r_2}{R_1}\right)^{2n} = \frac{(1-\varepsilon_1(\omega))(1-\varepsilon_2(\omega))}{(1+\varepsilon_1(\omega))(1+\varepsilon_2(\omega))} \quad (2)$$

where $\varepsilon_1$ and $\varepsilon_2$ are the permittivity of the shell and the core, respectively. Equation (1) indicates that the charge density on the inner surface of the cavity and outer surface of the core can be constructed from linear combinations of the surface charge densities of its two primitive components. Analogous to molecular orbital theory, the interaction of the spoof surface plasmons gives rise to a bonding mode with lower energy level corresponding the symmetric coupling and an antibonding mode with higher energy level corresponding to the asymmetric coupling, as shown in Fig. 1(a). In contrast to metal nanoshells in optical frequencies where surface charges exist in both metal and dielectric, the interaction of spoof SPPs in the PEC core and shell can only occur within the dielectric gap in between. Therefore, spoof plasmon modes of PEC structures can be tuned over a smaller range of kinetic energies than plasmon modes in metallic nanostructures. Fig. 1(b) displays the spectral positions of eigenmodes (i. e. magnetic dipolar mode, electric dipolar mode, and quadrupole mode) in terms of the gap size between the cavity and the core. Here, the radius of spiral core is kept as $R_1$=10mm, and the depth of the corrugations on the inner surface of the cavity is fixed as $R_2 - r_2 = 6.42 mm$. We find that as the gap size increases, the splitting between the bonding and antibonding modes decreases, due to the decrease of coupling between the charges on the surfaces of the core and the cavity. As the gap size approaches infinity, the eigenmodes will approach those of the individual primitive structures.

### 3. Experimental Verification

The spoof SPP hybridization picture illustrated with the example of spiral core-shell structures can also be used to understand the plasmon resonance of more complex spoof SPP structures such as dimers. The dimer structure has been considered as an important prototypical system that exhibits large electromagnetic field enhancement at plasmon resonances. Therefore, in what follows we experimentally study the hybridization of spoof SPPs and the underlying physics in the dimer structure in microwave frequencies. The schematic of the dimer structure is shown in Fig. 2(a), which consists of two identical metallic spiral structures separated by gap $g$=0.1 mm. Each spiral structure is composed of a small disk of radius $r$=1mm surrounded by six spiral arms with radius $R$=10mm. The arms have a width 0.5mm, and the spacing between neighbouring arms is 1.5mm, giving the period of the spiral $d$=2mm (as measured at the end of the arm). These two 0.035mm ultrathin metallic spiral disks are printed on a 0.5 mm-thick dielectric substrate with the relative permittivity 2.65 and loss tangent 0.002. Fig. 2(b) shows the photograph of the metallic spiral disks fabricated with a standard printed circuit board manufacturing process. To investigate the coupling between two closely spaced spiral disks and demonstrate how the plasmon modes of individual spiral disks interact and form new hybridized modes, we perform both numerical and experimental studies under dipole illuminations at the centre of the gap polarising horizontally and perpendicularly, as shown in Fig. 2(b). Finite-element simulations are carried out using a commercial software, CST Microwave Studio, where the metal is treated using the PEC approximation at microwave frequencies. In the measurement, a monopole probe is placed 1.5mm above the structure, and connected with the vector network analyzer (Agilent N5230C), which measures the reflection from the structure.

### 3.1. Hybridization of Spoof LSP Modes

The simulated and measured S11 spectra for the spiral dimer under two different polarized excitations are shown in Fig. 3, with the nearfield $E_z$ field distribution given in the right column. The dipolar modes of each spiral are hybridized to form plasmon orbitals, leading to new net dipole moments which govern the far-field optical properties [28, 29]. Here we focus our attention on the lowest momentums, i. e. dipolar modes and magnetic modes; higher-order multipolar modes are not discussed. As shown in Fig. 3(a), as the monopole probe is oriented perpendicular to the dimer, the dipolar modes of each spiral are excited, orienting perpendicularly to the dimer. Interesting, due to the non-mirror-symmetric geometry of the dimer, a degenerated mode with dipole oriented parallel to the dimer is formed (mode I), with a binding plasmon orbital at lowest energy (around 2.03 GHz). A standard bonding mode is formed at a higher frequency (2.2 GHz) with the dipoles in the two spirals in phase, oriented in the same direction as the external excitation (mode II). As demonstrated in previous work [22, 24], a magnetic dipolar mode can be excited on individual metallic spiral structures, where the circulating $E_z$ field varies between negative and positive to generate a closed current loop and a magnetic momentum. In Fig. 3(b), we can see that the magnetic plasmon modes on two spirals are anti-symmetrically coupled (antibonding) at 3.45 GHz (mode III). The phase is uniform in the azimuthal direction of each spiral and the plasmons on two spirals are oscillating out-of-phase. When the external dipolar excitation is parallel to the dimer, similarly, a degenerated mode with out-of-phase dipole moments on two spirals perpendicular to the dimer is observed at 2.15 GHz (mode IV), as shown in Fig. 3(c). The destructive interference of the two spiral plasmons forms an antibonding mode at around 2.25 GHz (mode V). At higher frequencies, the magnetic modes on individual spirals are both anti-symmetrically coupled (antibonding mode VI) and symmetrically coupled (bonding mode VII) with out-of-phase and in-phase oscillations on the two spirals. The simulation results agree well with the experiment in predicting the spectral positions of the resonance modes, while deviations are visible in both the magnitude and linewidth of the resonances, due to the difference between the monopole probe in the simulation and the measurement.

### 3.2. Extremely Large Field Enhancements

Similar to the core-shell system as discussed above, the interaction of spoof SPPs in the dimer structure can be engineered by the separation between the constitutive components. To study the splitting of bonding and antibonding dimer plasmons in terms of the separation and demonstrate how the spoof dipole resonance of two interacting spiral disks can be used to enhance the electric field in the centre of the gap, we change the mutual separation between the spirals and explore the variation of the resonance spectrum and the field enhancement at the gap, as shown in Fig. 4. When the gap size is gradually reduced from 0.9mm to 0.1mm, the the splitting of bonding and antibonding modes increases as the mutual coupling increases, and this interaction induces a red shift in the lowest mode of plasmon resonance (mode I), as confirmed by the simulation and measurement. Numerical simulations reveal the expected behaviour of strongly localized fields in the gap, which increases as the gap size decreases (insets of Fig. 4(a)). As given in Fig. 4(b), the plasmon resonance for mode I shifts predictably towards the red, and the field enhancement grows as the gap dimension decreases steadily, reaching up to 5500 at the gap size 0.1mm.

### 3.3. Radiative Effects

To achieve even larger enhancements of field, we may wish to further decrease the gap size, which is ultimately limited by the fabrication technique. An alternative approach is to increase the size of the spiral disks while keeping the separation unchanged, a way to reduce the relative gap size. However, as the size of the spiral structure grows, the radiative damping plays an increasingly important part, leading to the reduction of field enhancements. Therefore, it is vital to take into account the trade-off between the interaction and the radiative reaction, and find the dimension parameters which give the optimum performance in light harvesting. Here, we investigate this problem with numerical simulations by comparing the extinction cross sections for spiral dimers with different disk radii from 9mm to 13mm while the gap size remains 0.1mm. Three resonance peaks can be observed at $R$=9mm in Fig. 5(a). The increase of the disk dimension not only leads to the distinct red shift of all the resonance peaks as expected, but also results in the change of the resonance modes and the enhancement of electromagnetic field at the gap. As displayed in Fig. 5(b), at the first resonance peak (i.e. the lowest resonance frequency), the field enhancement grows with the increase of the disk dimension and reach the maximum at $R$=11mm. As we further increase the disk size, the contribution of the radiation loss will surpass that from the reduction of relative gap dimension, leading to the decrease of the field enhancement. The gap field enhancement at the second resonance peak corresponding to a quadruple mode is monotonically increasing with the disk size from 9 mm to 11 mm. At the third resonance peak, the field enhancement slightly decreases as the radius change from 9 mm to 10 mm, and then steadily increases with *R*. Notably, as the disk radius reaches 13mm, the field enhancements of higher-order modes (the second and third resonance peaks) exceed those of the dipolar resonance. This phenomenon is quite different from that observed for nanoparticle dimer at optical frequencies, where the maximum field enhancements generally occur at the lowest resonance peak [10, 30, 31].

**3.4. Spatial Nonlocality**

Besides very small gaps, fabricating very small corrugations (or very large corrugation number) is also challenging in experiments of spiral dimers supporting spoof SPPs. Here, we consider the effect of spatial nonlocality resulting from the finite size of the surface textures and find how the spatial nonlocality affects the maximum field enhancement. As shown in Fig. 6, the resonance frequencies of the dipolar modes and corresponding field enhancement are calculated in terms of the number of corrugations in each spiral disk. The resonance shows a red shift as the number of spiral arms increases, as denoted by the black curve. The field enhancement grows as the spatial nonlocality is reduced, and will gradually converge for arm number larger than 16.

## 4. Conclusion

We have shown that an intuitive hybridization picture analogous to what has been previously established for plasmonic nanostructures can be applied for describing and understanding the spoof SPP modes in composite metal structures at low frequencies (from microwave to terahertz spectra). The spoof plasmon hybridization scheme is theoretically demonstrated using the example of a 2D textured PEC core-shell structure, representing the simplest coupling system, and further investigated experimentally in a more complex metal dimer structure which support spoof SPPs at microwave frequencies. We have shown that the interaction of the spoof SPPs gives rise to bonding modes and antibonding modes. The spectral positions of collective modes and coupling strength depend on the geometries and spectra of the primitive structures, allowing for the manipulation of hybridization to produce enormous field enhancement. Practical issues, including spatial nonlocality arising from the finite size of the surface textures and the radiation loss particularly in large structures, have been considered to find the optimized geometrical parameters for which the field enhancement is maximized.

**Acknowledgement**

The authors wish to thank Professor John Pendry for fruitful discussions. This work was supported in part by Nanyang Technological University Start-up Grants, Singapore Ministry of Education (MOE) under Grant No RG72/15 and Grant No. MOE2015-T2-1-145, and in part by the National Science Foundation of China (61571117, 61171024, 61171026, 61138001), 111 Project (111-2-05), National Instrumentation Program (2013YQ200647), Scientific Research Foundation of Graduate School of Southeast University (YBJJ1436), and Program for Postgraduate Research Innovation in University of Jiangsu Province (3204004910). S.A.M. acknowledges the EPSRC (EP/L 204926/1), the Royal Society, and the Lee-Lucas Chair in Physics.

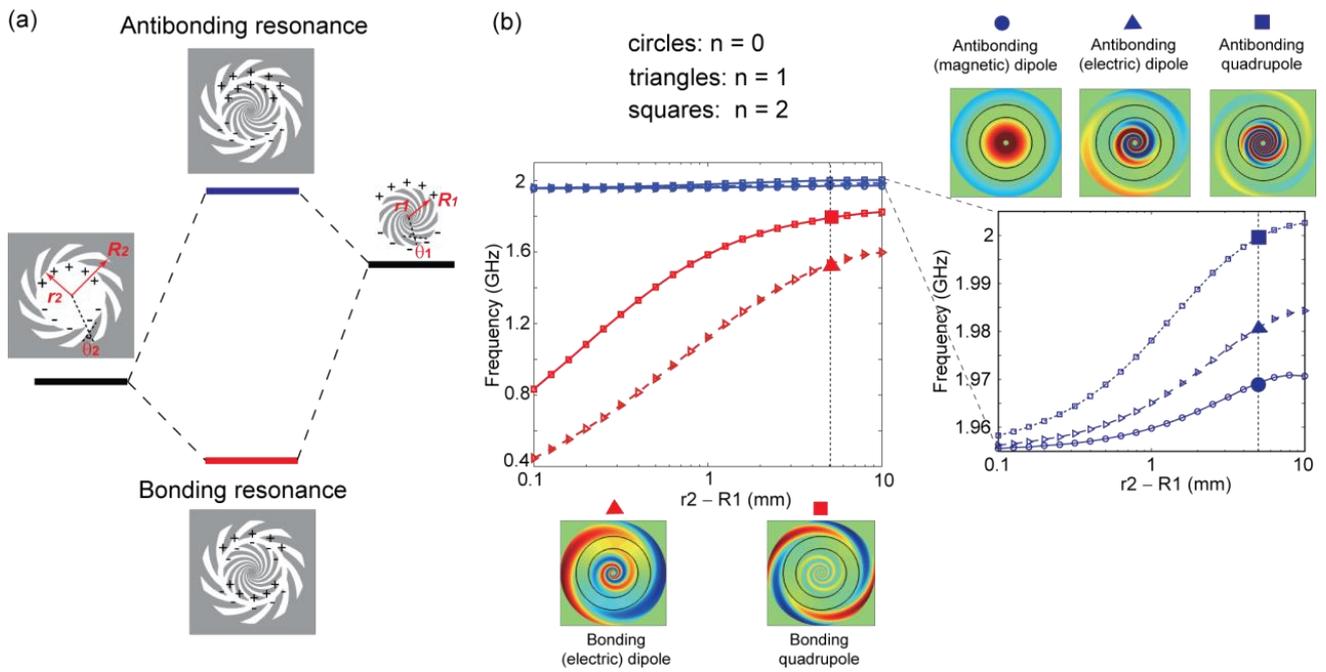

Figure 1. (a) An energy-level diagram describing the spoof plasmon hybridization in PEC structures comprising a PEC cavity with spiral corrugations and a spiral PEC core. The interaction between the cavity and core spoof-plasmons results in an antibonding and a bonding spoof plasmon mode. (b) Spectral positions of eigenmodes in terms of the gap size between the cavity and the core. The electric field distributions of each modes at the gap size 5mm are plotted in the insets.

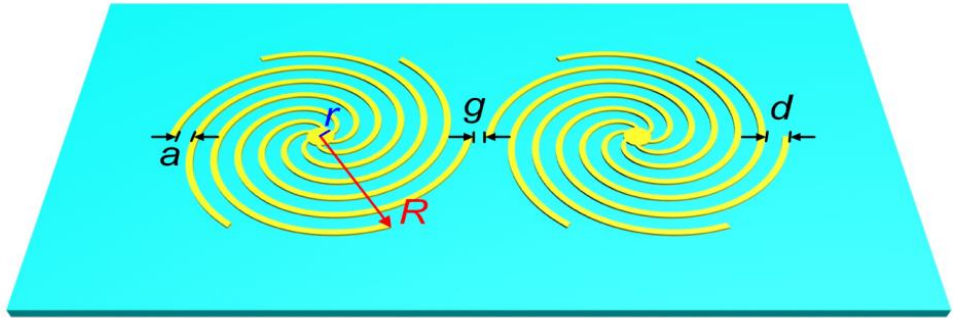

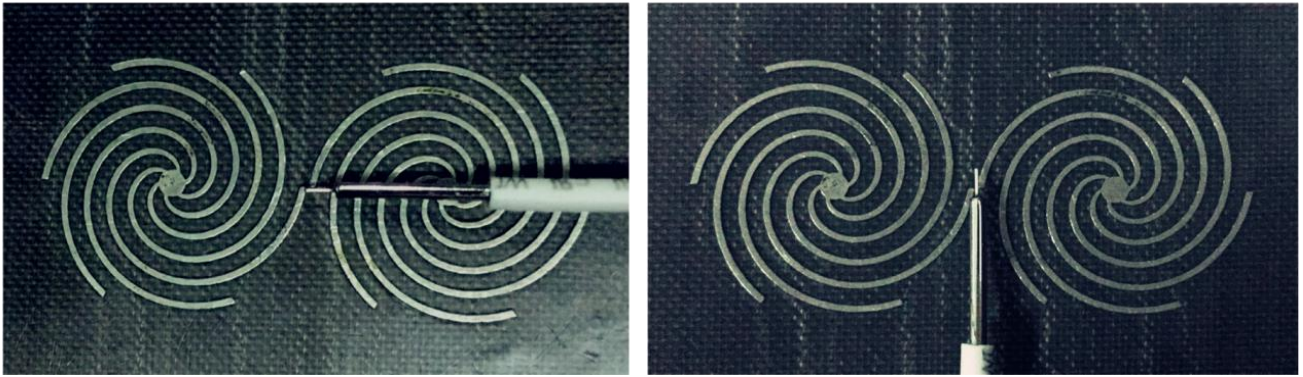

Figure 2: (a) Schematic of the closely spaced spiral structures. (b) Photograph of adjacent spiral structures under horizontal (left) and perpendicular (right) dipole excitations at the gap between the spirals. The gap between two spirals is $g$=0.1 mm. Each spiral structure is composed of a small disk of radius r=1mm surrounded by six spiral arms with radius $R$=10mm. The period of the spiral arms is $d$=2mm, and the spacing between neighbouring arms is $a$=1.5mm. These two 0.035mm ultrathin metallic spiral disks are printed on a 0.5 mm-thick dielectric substrate with the relative permittivity 2.65 and loss tangent 0.002.

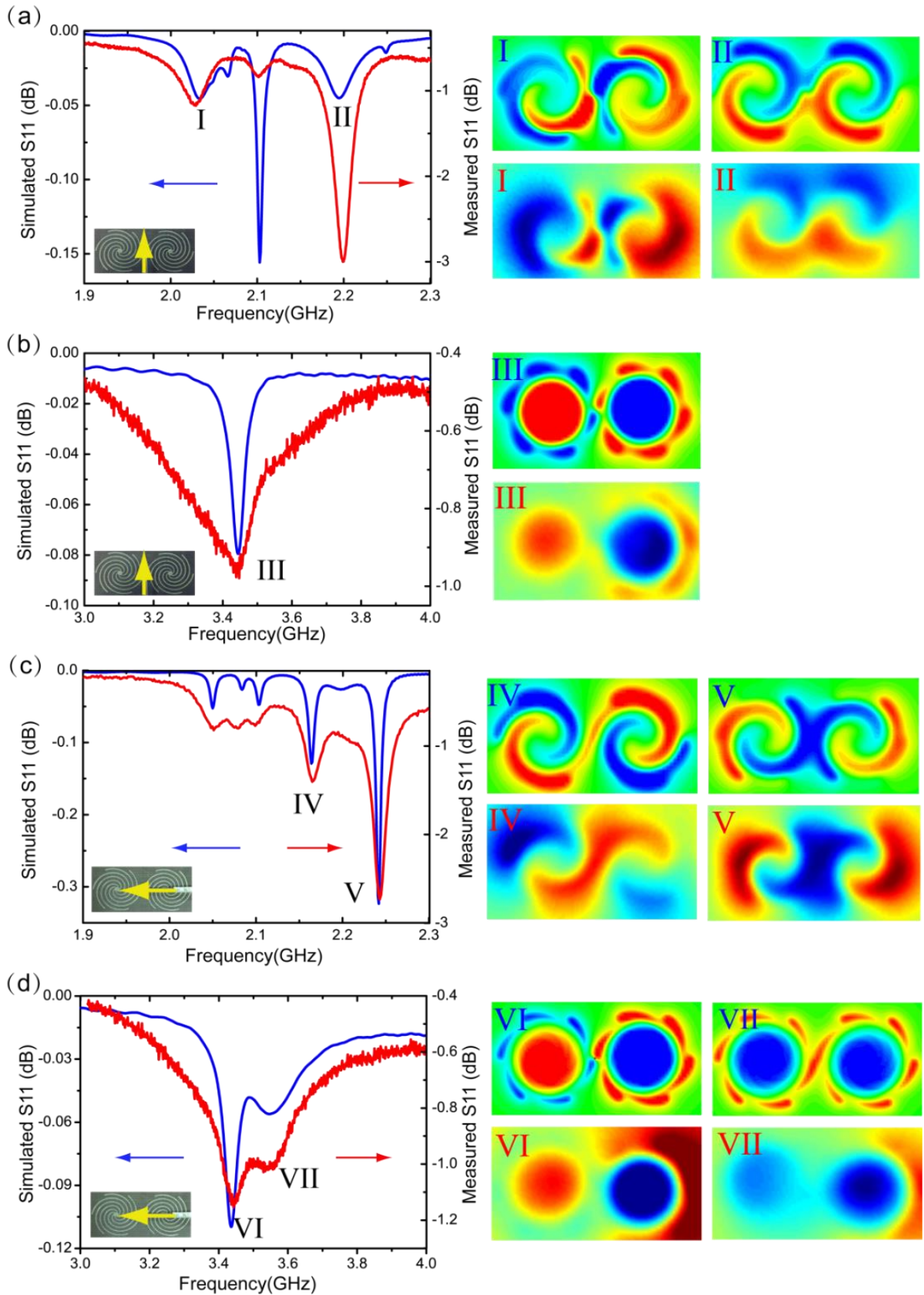

Figure 3. Simulated and measured reflection spectra under (a,b) horizontal and (c,d) perpendicular dipole excitations (left column), and the corresponding simulated and measured near field distributions for different resonance modes (right column).

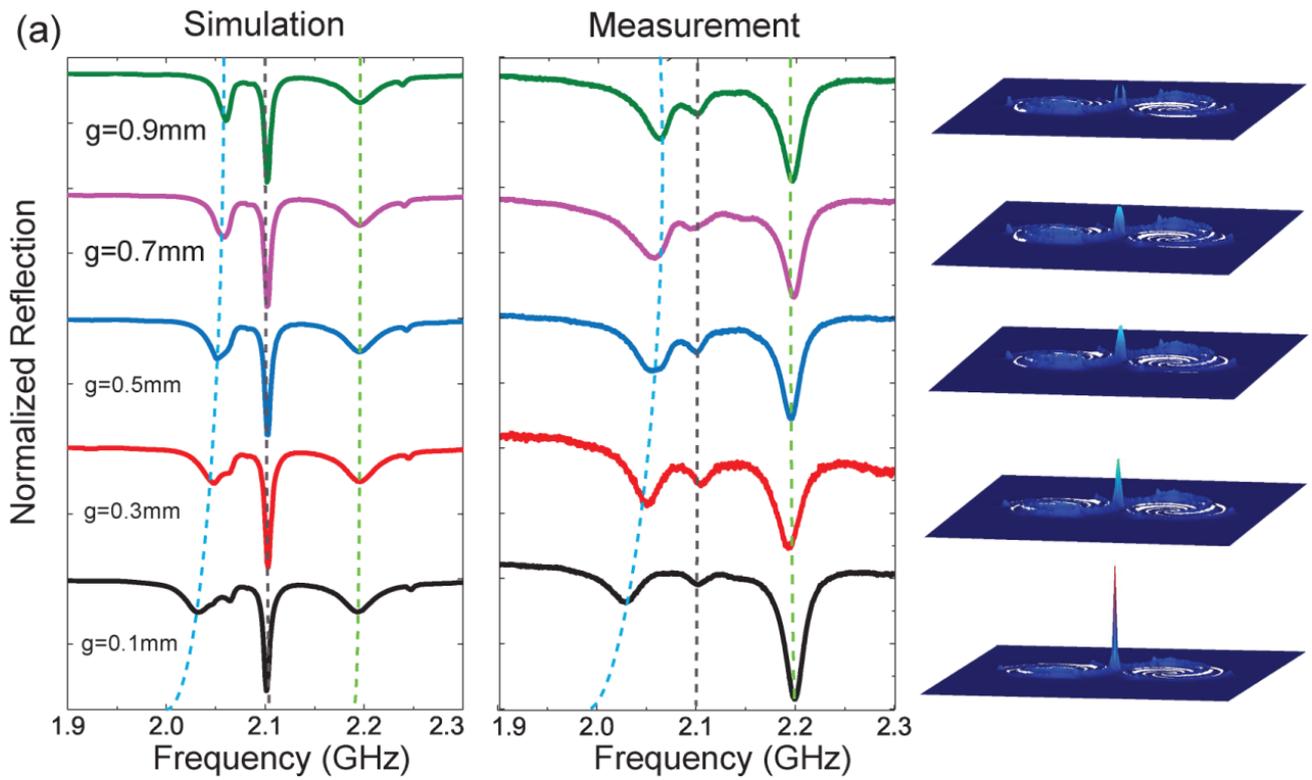

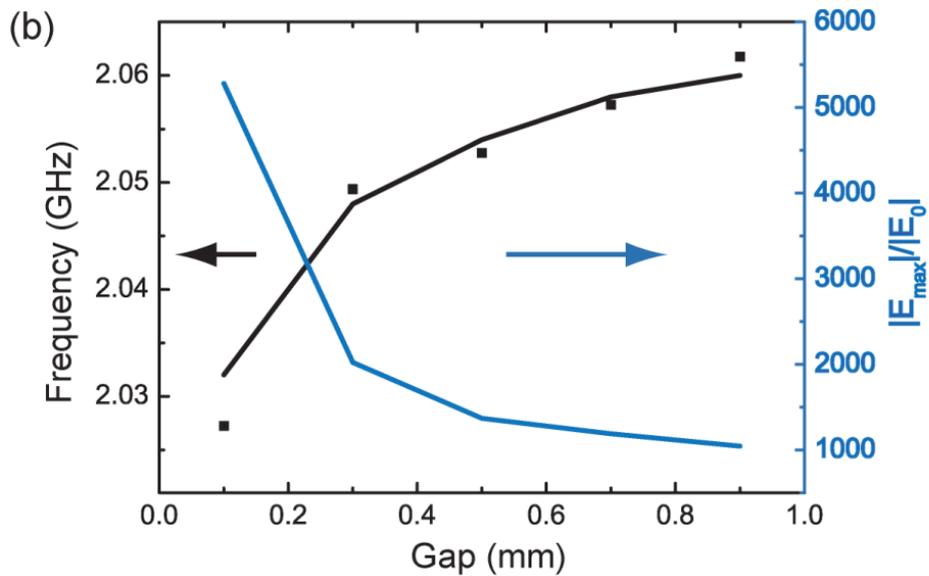

Figure 4. (a) Simulation and measurement of reflection spectra for adjacent spiral structures with varying gap size from 0.9mm to 0.1mm. The insets show the corresponding electric field enhancements at the center of the gap. (b) The resonance frequency of mode I (black) and the maximum field enhancement (blue) as a function of the gap size. Here the solid curves correspond to the simulation results and the dots denote the measurement result.

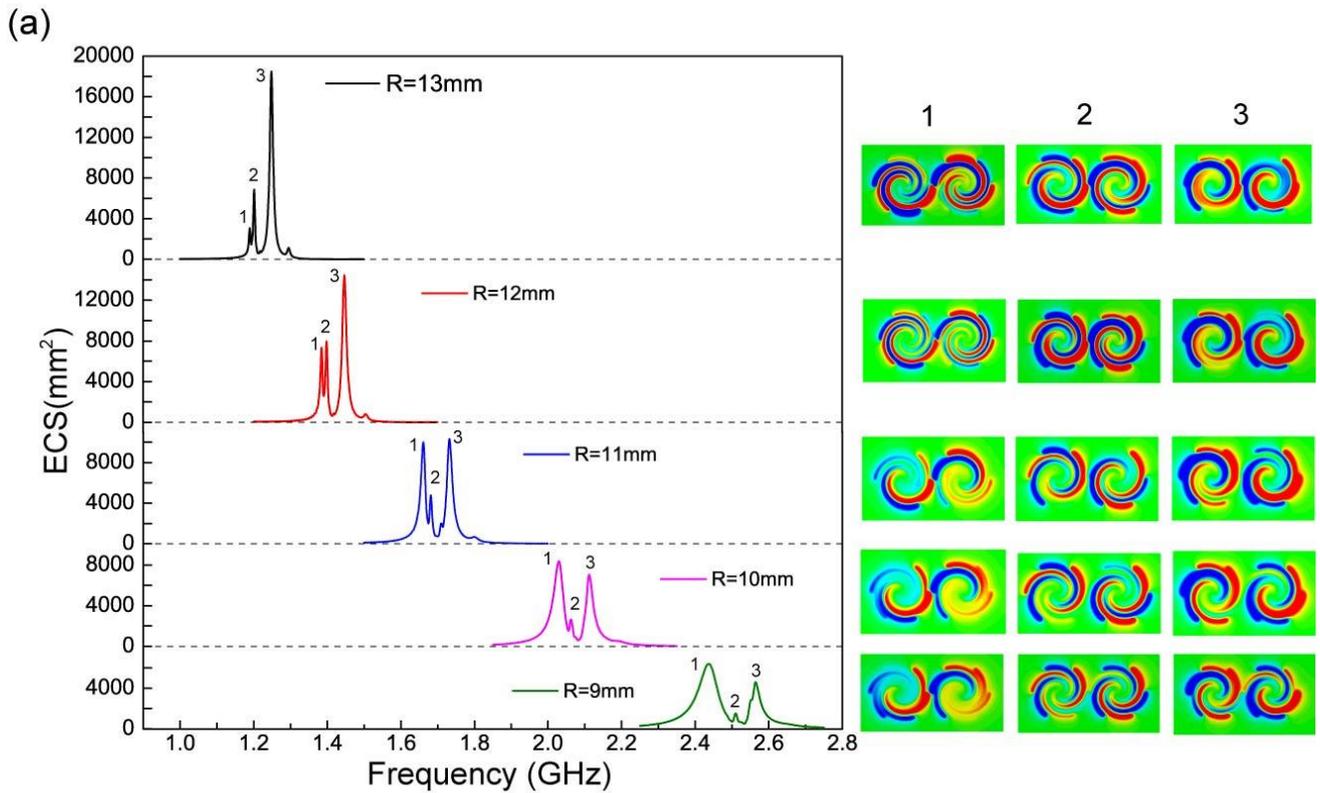

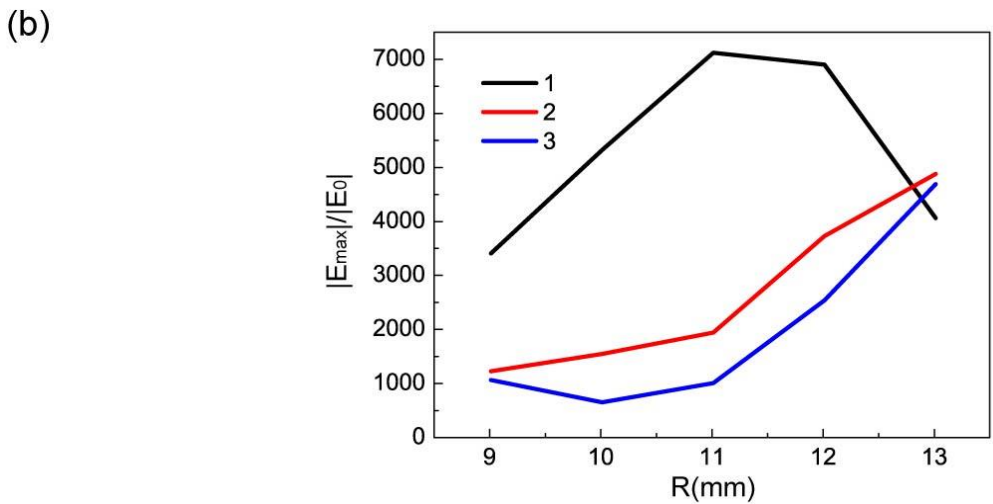

Figure 5. (a) Simulated extinction cross section for spiral dimers with different disk radii varying from R=9mm to R=13mm and the same gap size g=0.1mm. The insets in the right column show the corresponding field distributions for different resonance peaks. (b) The maximum field enhancement at different resonance peaks as a function of the spiral disk radius.

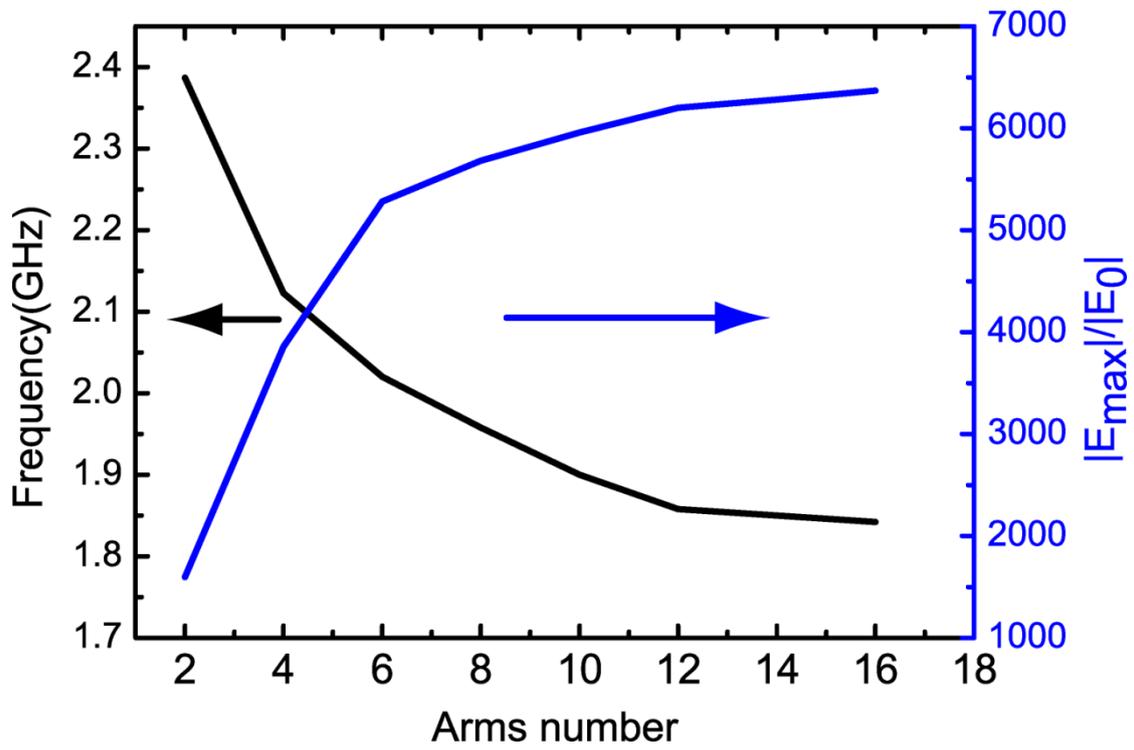

Figure 6 Simulated dipolar resonance frequency (black) and the field enhancement at the gap (blue) as a function of the number of spiral arms.

# Supporting Information

**Enormous Field Enhancement in Hybrid Spoof Plasmonic System**

*Jingjing Zhang, Zhen Liao, Yu Luo \*, Xiaopeng Shen, Stefan A. Maier \*, and Tie Jun Cui \**

\* To whom correspondence should be addressed;
E-mail: luoyu@ntu.edu.sg, s.maier@imperial.ac.uk, tjcui@seu.edu.cn

We consider the corrugated structures with spiral grooves. The intersection angle between each spiral and the radial direction is defined as $\theta$. The effective permittivity and permeability of the corresponding metamaterial cover can be written as

$$\bar{\bar{\varepsilon}}^{-1} = \frac{a}{2n_g^2 d}\begin{bmatrix} 1-\cos 2\theta & -\sin 2\theta \\ -\sin 2\theta & 1+\cos 2\theta \end{bmatrix}, \quad \mu_z^{-1} = \frac{d}{a}. \tag{S1}$$

The magnetic field in each region can be written as

$$H_z = \begin{cases} \sum_{n=-\infty}^{\infty}\left[ f_n J_0\left(\frac{n_g k_0}{\cos\theta_2}\rho\right) + g_n Y_0\left(\frac{n_g k_0}{\cos\theta_2}\rho\right)\right] e^{in(\varphi-\tan\theta_2 \ln\rho)}, & R_2 \geq \rho \geq r_2 \\ \sum_{n=-\infty}^{\infty}\left[ a_n J_n(k_0\rho) + b_n H_n^{(1)}(k_0\rho)\right] e^{in\varphi}, & r_2 \geq \rho \geq R_1 \\ \sum_{n=-\infty}^{\infty}\left[ c_n J_0\left(\frac{n_g k_0}{\cos\theta_1}\rho\right) + d_n Y_0\left(\frac{n_g k_0}{\cos\theta_1}\rho\right)\right] e^{in(\varphi-\tan\theta_1 \ln\rho)}, & R_1 \geq \rho \geq r_1 \end{cases} \tag{S2}$$

The electric fields can be simply obtained by $\bar{E} = (i/\omega)\bar{\bar{\varepsilon}}\cdot\nabla\times\bar{H}$. Then, by applying the boundary conditions we have the following equations:

$$a_n J_n(k_0 r_2) + b_n H_n^{(1)}(k_0 r_2) = r_2^{-in\tan\theta_2}\left[ f_n J_0\left(\frac{n_g k_0 r_2}{\cos\theta_2}\right) + g_n Y_0\left(\frac{n_g k_0 r_2}{\cos\theta_2}\right)\right] \tag{S3}$$

$$a_n J_n'(k_0 r_2) + b_n H_n^{(1)\prime}(k_0 r_2) = -\frac{a\cos\theta_2}{n_g d} r_2^{-in\tan\theta_2}\left[ f_n J_1\left(\frac{n_g k_0 r_2}{\cos\theta_2}\right) + g_n Y_1\left(\frac{n_g k_0 r_2}{\cos\theta_2}\right)\right] \tag{S4}$$

$$f_n J_1\left(\frac{n_g k_0 R_2}{\cos\theta_2}\right) + g_n Y_1\left(\frac{n_g k_0 R_2}{\cos\theta_2}\right) = 0 \tag{S5}$$

$$a_n J_n(k_0 R_1) + b_n H_n^{(1)}(k_0 R_1) = R_1^{-in\tan\theta_1}\left[c_n J_0\left(\frac{n_g k_0 R_1}{\cos\theta_1}\right) + d_n Y_0\left(\frac{n_g k_0 R_1}{\cos\theta_1}\right)\right] \quad (S6)$$

$$a_n J_n'(k_0 R_1) + b_n H_n^{(1)'}(k_0 R_1) = -\frac{a\cos\theta_1}{n_g d} R_1^{-in\tan\theta_1}\left[c_n J_1\left(\frac{n_g k_0 R_1}{\cos\theta_1}\right) + d_n Y_1\left(\frac{n_g k_0 R_1}{\cos\theta_1}\right)\right] \quad (S7)$$

$$c_n J_1\left(\frac{n_g k_0 r_1}{\cos\theta_1}\right) + d_n Y_1\left(\frac{n_g k_0 r_1}{\cos\theta_1}\right) = 0. \quad (S8)$$

Eliminating $c_n$, $d_n$, $f_n$, and $g_n$ in the equations above yields

$$\frac{a_n J_n'(k_0 r_2) + b_n H_n^{(1)'}(k_0 r_2)}{a_n J_n(k_0 r_2) + b_n H_n^{(1)}(k_0 r_2)} = -\frac{a\beta\cos\theta_2}{n_g d}, \quad (S9)$$

$$\frac{a_n J_n'(k_0 R_1) + b_n H_n^{(1)'}(k_0 R_1)}{a_n J_n(k_0 R_1) + b_n H_n^{(1)}(k_0 R_1)} = \frac{a\alpha\cos\theta_1}{n_g d}, \quad (S10)$$

where we have defined two coefficients

$$\alpha = \frac{J_1(n_g k_0 r_1/\cos\theta_1)Y_1(n_g k_0 R_1/\cos\theta_1) - J_1(n_g k_0 R_1/\cos\theta_1)Y_1(n_g k_0 r_1/\cos\theta_1)}{J_0(n_g k_0 R_1/\cos\theta_1)Y_1(n_g k_0 r_1/\cos\theta_1) - J_1(n_g k_0 r_1/\cos\theta_1)Y_0(n_g k_0 R_1/\cos\theta_1)}, \quad (S11)$$

$$\beta = \frac{J_1(n_g k_0 r_2/\cos\theta_2)Y_1(n_g k_0 R_2/\cos\theta_2) - J_1(n_g k_0 R_2/\cos\theta_2)Y_1(n_g k_0 r_2/\cos\theta_2)}{J_0(n_g k_0 r_2/\cos\theta_2)Y_1(n_g k_0 R_2/\cos\theta_2) - J_1(n_g k_0 R_2/\cos\theta_2)Y_0(n_g k_0 r_2/\cos\theta_2)}. \quad (S12)$$

Further simplifying Equation (S9) and (S10) leads to the resonance condition

$$\frac{J_n'(k_0 R_1) - \frac{a\alpha\cos\theta_1}{n_g d} J_n(k_0 R_1)}{H_n^{(1)'}(k_0 R_1) - \frac{a\alpha\cos\theta_1}{n_g d} H_n^{(1)}(k_0 R_1)} = \frac{J_n'(k_0 r_2) + \frac{a\beta\cos\theta_2}{n_g d} J_n(k_0 r_2)}{H_n^{(1)'}(k_0 r_2) + \frac{a\beta\cos\theta_2}{n_g d} H_n^{(1)}(k_0 r_2)}. \quad (S13)$$

Since we are interested in subwavelength particles where $k_0 a_2 < k_0 b_1 \ll 1$, Equation (S13) can be rewritten as

$$\left(\frac{r_2}{R_1}\right)^{2n} = \frac{\left(1 - \frac{k_0 a R_1 \cos\theta_1 \alpha}{n n_g d}\right)\left(1 - \frac{k_0 a r_2 \cos\theta_2 \beta}{n n_g d}\right)}{\left(1 + \frac{k_0 a R_1 \cos\theta_1 \alpha}{n n_g d}\right)\left(1 + \frac{k_0 a r_2 \cos\theta_2 \beta}{n n_g d}\right)}. \quad (S14)$$

Defining

$$\varepsilon_1 = \frac{k_0 a R_1 \cos\theta_1 \alpha}{n n_g d} \text{ and } \varepsilon_2 = \frac{k_0 a r_2 \cos\theta_2 \beta}{n n_g d}, \tag{S15}$$

Eq. (S14) reduces to

$$\left(\frac{r_2}{R_1}\right)^{2n} = \frac{(1-\varepsilon_1)(1-\varepsilon_2)}{(1+\varepsilon_1)(1+\varepsilon_2)}, \tag{S16}$$

which is the resonance condition of a core-shell structure.